\batchmode
\makeatletter
\def\input@path{{C:/Users/Xiang/Dropbox/HeteroM/Methods/CDE_ms/}}
\makeatother
\documentclass[11pt,english]{article}\usepackage[]{graphicx}\usepackage[]{color}
\makeatletter
\def\maxwidth{ %
  \ifdim\Gin@nat@width>\linewidth
    \linewidth
  \else
    \Gin@nat@width
  \fi
}
\makeatother

\definecolor{fgcolor}{rgb}{0.345, 0.345, 0.345}

\usepackage{framed}
\makeatletter
 {\par\unskip\endMakeFramed%
 \at@end@of@kframe}
\makeatother

\definecolor{shadecolor}{rgb}{.97, .97, .97}
\definecolor{messagecolor}{rgb}{0, 0, 0}
\definecolor{warningcolor}{rgb}{1, 0, 1}
\definecolor{errorcolor}{rgb}{1, 0, 0}

\usepackage{alltt}
\usepackage[latin9]{inputenc}
\usepackage[letterpaper]{geometry}
\usepackage{mathtools}
\usepackage{amsmath}
\usepackage{amsthm}
\usepackage{amssymb}
\usepackage{graphicx}
\usepackage{setspace}
\usepackage[authoryear]{natbib}
\makeatletter
\theoremstyle{plain}
\newtheorem{prop}{\protect\propositionname}

\usepackage{amsfonts}
\usepackage{amsthm}
\usepackage{fancyhdr}
\usepackage{lastpage}
\usepackage{extramarks}
\usepackage{chngpage}
\usepackage{soul}
\usepackage{color}\usepackage{graphics}
\usepackage{lscape}
\usepackage{url}

\usepackage{bm}
\usepackage{enumerate}
\usepackage{verbatim}
\usepackage{babel}
\usepackage{titlesec}

\def\ci{\perp\!\!\!\perp}

\usepackage{pifont}

\titleformat{\subsection}
  {\normalfont\large\bfseries}{\thesubsection}{1em}{}

\usepackage{pifont}
\usepackage{pgf}
\usepackage{tikz}\usetikzlibrary{positioning}
\usetikzlibrary{arrows}

\usepackage{footmisc}

\allowdisplaybreaks

\usepackage{sectsty}

\makeatother

\usepackage{babel}
\providecommand{\propositionname}{Proposition}
\IfFileExists{upquote.sty}{\usepackage{upquote}}{}
\begin{document}
\title{\vspace{-20pt}
Some Doubly and Multiply Robust Estimators of Controlled Direct Effects\thanks{Direct all correspondence to Xiang Zhou, Department of Sociology,
Harvard University, 33 Kirkland Street, Cambridge MA 02138; email:
xiang\_zhou@fas.harvard.edu. The author thanks Aleksei Opacic for
helpful comments.}}
\author{Xiang Zhou}
\date{November 18, 2020}
\maketitle
\begin{abstract}
\noindent \sloppy This letter introduces several doubly, triply,
and quadruply robust estimators of the controlled direct effect. Among
them, the triply and quadruply robust estimators are locally semiparametric
efficient, and well suited to the use of data-adaptive methods for
estimating their nuisance functions.
\end{abstract}

\section{Introduction}

Over the past decade, causal mediation analysis has grown popular
in social and biomedical sciences. A common approach to assessing
causal mediation involves decomposing the total effect of a treatment
on an outcome into the so-called natural direct and indirect effects
(NDE and NIE; \citealt{robins1992direct,pearl2001direct}). The NDE
and NIE, however, are not nonparametrically identified in the presence
of posttreatment confounders, i.e., when confounders of the mediator-outcome
relationship may be affected by the treatment itself (\citealt{avin2005identifiability,vanderweele2009conceptual}).
In such cases, researchers have often focused on estimating the controlled
direct effect (CDE), a quantity that measures the effect of treatment
when a mediator is fixed at a given value for all units (\citealt{pearl2001direct}).
Thus a nonzero CDE implies that the effect of treatment on the outcome
does not operate exclusively through the mediator of interest. Unlike
the NDE and NIE, the CDE is identified provided that all confounders
for the treatment-outcome relationship and for the mediator-outcome
relationship are observed, even if some of the mediator-outcome confounders
are affected by the treatment itself.

Estimators of the CDE typically rely on correct specification of (at
least) two nuisance functions about the conditional means/densities
of the treatment, mediator, outcome, or posttreatment confounders.
For example, the weighting estimator proposed by \citet{vanderweele2009marginal}
involves fitting two propensity score models, one for the treatment
and one for the mediator, and the sequential g-estimator proposed
by \citet{vansteelandt2009estimating} involves fitting two outcome
models, one for the observed outcome given all its antecedent variables
and one for a \textquotedblleft demediated\textquotedblright{} outcome
given pretreatment confounders and the treatment (see also \citealt{joffe2009related}).
To alleviate bias due to model misspecification, \citet{goetgeluk2008estimation}
proposed a doubly robust estimator of the CDE that depends on correct
specification of (a) a model for treatment assignment, and (b) either
an outcome model or a propensity score model for the mediator.

The causal structure underlying identification and estimation of the
CDE is akin to that of estimating treatment effects with longitudinal
data in the presence of time-varying confounders (\citealt{robins1999marginal}).
For the latter problem, \citet{bang2005doubly} have proposed a doubly
robust estimator for the mean of a potential outcome that depends
on correct specification of either (a) propensity score models for
treatment status at all time points or (b) models for an iteratively
imputed outcome at all time points. In a recent paper, \citet{rotnitzky2017multiply}
point out that the Bang-Robins estimator is actually ``multiply robust''
because it is consistent whenever the first $k$ propensity score
models and the last $K-k$ ``outcome models'' are correctly specified,
where $0\leq k\leq K$, and $K$ is the number of time points (see
also \citealt{molina2017multiple}). Moreover, these authors show
that the Bang-Robins procedure can be further improved with a $2^{K}$-robust
estimator that requires correct specification of either the propensity
score model or the outcome model at each time point.\footnote{In a separate strand of literature, the term ``multiple robustness''
has been used to characterize a class of estimators for the mean of
incomplete data in a cross-sectional setting that are consistent if
one of several models for the propensity score or one of several models
for the outcome is correctly specified (e.g., \citealt{han2013estimation}).
Following \citet{molina2017multiple} and \citet{rotnitzky2017multiply},
we use ``multiple robustness'' to characterize estimators that require
modeling \textit{multiple parts of the observed data likelihood }and
are consistent if one of several sets of the corresponding models
are correctly specified.}

Capitalizing on the above work, this letter introduces a set of doubly
robust, triply robust, and quadruply robust estimators of the CDE,
which, to the best of the author's knowledge, are new to the causal
mediation literature. While some of these estimators ($\hat{\psi}_{am}^{\textup{tr}_{1}}$
and $\hat{\psi}_{am}^{\textup{qr}}$; see Section \ref{sec:Multiply-Robust})
are closely related to those proposed in \citet{bang2005doubly} and
\citet{rotnitzky2017multiply} for estimating time-varying treatment
effects, the rest of them have not been discussed elsewhere. These
estimators all involve estimating more than two nuisance functions;
yet, under suitable regularity conditions, they are consistent and
asymptotically normal (CAN) when only two of these nuisance functions
are correctly specified and their estimates are $\sqrt{n}$-consistent.
The triply and quadruply robust estimators are locally efficient,
i.e., when all of the nuisance functions are correctly specified,
they attain the semiparametric efficiency bound in the nonparametric
model over observed data. Moreover, their estimating equations are
Neyman orthogonal, encouraging the use of machine learning methods
and cross-fitting to estimate the nuisance functions (\citealt{zheng2011cross,chernozhukov2018double}),
in which case estimates of the CDE are semiparametric efficient when
estimates of the nuisance functions, for example, all converge at
faster-than-$n^{-1/4}$ rates.

\section{Notation, Assumptions, and Identification\label{subsec:Notation-Assumptions}}

Let $A$ denote treatment, $M$ the mediator, $Y$ the observed outcome,
and $Y(a,m)$ the potential outcome under treatment status $a$ and
mediator value $m$. We focus on the simple setting where the treatment
$A$ and the mediator $M$ are both discrete with finite support.
In addition, we denote by $X$ a vector of pretreatment variables
that may confound the causal effect of $(A,M)$ on $Y$, and denote
by $Z$ a vector of posttreatment variables that may confound the
causal effect of $M$ on $Y$. Note that the posttreatment confounders
$Z$ may themselves be affected by the treatment.

The controlled direct effect (CDE) is defined as the average effect
of switching treatment status from $a'$ to $a$ while fixing the
mediator at a given level $m$:
\[
\textup{CDE}(a,a',m)=\mathbb{E}[Y(a,m)-Y(a',m)]
\]
By definition, the CDE is identified when the expected potential outcome
$\mathbb{E}[Y(a,m)]$ is identified for any $a$ and $m$. Thus, we
focus on the latter estimand throughout the paper and denote it as
$\psi_{am}$. Since it is the expected potential outcome when both
the treatment and the mediator are ``controlled'' at given values,
we may refer to it as the controlled response function (CRF). The
CRF can also be used to construct other estimands such as the controlled
mediator effect $\textup{CME}(a,m,m')=\psi_{am}-\psi_{am'}$ (\citealt{zheng2015causal})
and the treatment-mediator interaction effect $(\psi_{am}-\psi_{am'})-(\psi_{a'm}-\psi_{a'm'})$.

The CRF is identified under the assumptions of consistency, sequential
ignorability, and positivity:
\begin{enumerate}
\item consistency: for any unit, if $A=a$ and $M=m$, then $Y=Y(a,m)$;
\item sequential ignorability: $Y(a,m)\ci A|X,\forall a,m,$ and $Y(a,m)\ci M|X,A,Z$,
$\forall a,m$.
\item positivity: $p_{A|X}(a|x)>\epsilon>0$ and $p_{M|X,A,Z}(m|x,a,z)>\epsilon>0$
$\forall a,m$, $x\in\textup{supp}(X)$, and $z\in\textup{supp}(Z|X=x,A=a)$,
\end{enumerate}
where $p(\cdot)$ denotes a probability mass/density function. Under
assumptions 1-3, the CRF (and hence the CDE) can be identified via
Robins's (\citeyear{robins1986new}) g-computation formula:

\begin{equation}
\psi_{am}=\iiint ydP(y|x,a,z,m)dP(z|x,a)dP(x),\label{eq:identification}
\end{equation}
where $P(u|v)$ denotes the cumulative distribution function of $U$
given $V$.

\section{G-Computation, Imputation, and Weighting}

\sloppy Using the law of iterated expectations, equation \eqref{eq:identification}
can be written in several different forms, each of which points to
a different way of estimating the CRF:
\begin{align}
\psi_{am} & =\iint\mathbb{E}[Y|x,a,z,m]dP(z|x,a)dP(x)\quad\textup{(g-computation)}\label{eq:g-comp}\\
 & =\mathbb{E}_{X}\mathbb{E}_{Z|X,A=a}\mathbb{E}[Y|X,A,Z,M=m]\quad\textup{(pure imputation)}\label{eq:imputation}\\
 & =\ensuremath{\mathbb{E}\big[\frac{\mathbb{I}(A=a)\mathbb{E}[Y|X,A,Z,M=m]}{\Pr[A=a|X]}\big]}\quad\textup{(imputation-then-weighting)}\label{eq:imputation-weighting}\\
 & =\text{\ensuremath{\mathbb{E}\big[\frac{\mathbb{I}(A=a)\mathbb{I}(M=m)Y}{\Pr[A=a|X]\Pr[M=m|X,A,Z]}\big]}\quad\textup{(pure weighting)}}\label{eq:weighting}\\
 & =\ensuremath{\mathbb{E}_{X}\mathbb{E}\big[\frac{\mathbb{I}(M=m)Y}{\Pr[M=m|X,A,Z]}|X,A=a\big]\quad\textup{(weighting-then-imputation)}}\label{eq:weighting-then-imputation}
\end{align}

Equation \eqref{eq:g-comp} suggests a procedure akin to Robins's
g-computation algorithm: (1) fit a parametric model for the conditional
distribution of $Z$ given $X$ and $A$; (2) fit a parametric or
semiparametric model for the conditional mean of $Y$ given $X,A,Z$,
and $M$; and (3) evaluate the inner integral via Monte Carlo simulation
and the outer integral via the empirical distribution of $X$. In
the particular case where the models for $\mathbb{E}[Z|x,a]$ and
$\mathbb{E}[Y|x,a,z,m]$ are both linear, equation \eqref{eq:g-comp}
can be evaluated using a simple ``regression-with-residuals'' procedure
(\citealt{zhou2019regression}). Equation \eqref{eq:imputation} suggests
a ``pure imputation'' procedure: (1) fit a model for the conditional
mean of $Y$ given $X,A,Z$, and $M$ and obtain predicted values
for all units at $M=m$, $\hat{\mathbb{E}}[Y|X,A,Z,M=m]$; (2) fit
a model for the conditional mean of $\hat{\mathbb{E}}[Y|X,A,Z,M=m]$
given $X$ and $A$ and obtain its predicted values for all units
at $A=a$; (3) average these predicted values over all units. This
procedure is similar to the sequential g-estimator proposed in \citet{vansteelandt2009estimating}
and \citet{joffe2009related}. Equation \eqref{eq:imputation-weighting}
suggests an imputation-then-weighting procedure: (1) fit a model for
the conditional mean of $Y$ given $X,A,Z$, and $M$ and obtain predicted
values for all units at $M=m$, $\hat{\mathbb{E}}[Y|X,A,Z,M=m]$;
(2) fit a propensity score model for treatment status and obtain fitted
values $\widehat{\Pr}[A=a|X]$; (3) compute a weighted average of
the predicted outcomes $\hat{\mathbb{E}}[Y|X,A,Z,M=m]$ with inverse-probability
weights $\mathbb{I}(A=a)/\widehat{\Pr}[A=a|X]$.

Equation \eqref{eq:weighting} suggests a ``pure weighting'' estimator
(\citealt{vanderweele2009marginal}): (1) fit a propensity score model
for treatment status and obtain fitted values $\widehat{\Pr}[A=a|X]$;
(2) fit a propensity score model for the mediator and obtain fitted
values $\widehat{\Pr}[M=m|X,A,Z]$; and (3) compute a weighted average
of observed outcomes with inverse-probability weights $\mathbb{I}(A=a)\mathbb{I}(M=m)/(\widehat{\Pr}[A=a|X]\cdot\widehat{\Pr}[M=m|X,A,Z]\big)$.
Finally, equation \eqref{eq:weighting-then-imputation} suggests a
weighting-then-imputation procedure: (1) fit a propensity score model
for the mediator and obtain fitted values $\widehat{\Pr}[M=m|X,A,Z]$;
(2) fit a model for the conditional mean of the inverse-probability-weighted
outcome $\mathbb{I}(M=m)Y/\widehat{\Pr}[M=m|X,A,Z]$ given $X$ and
$A$ and obtain predicted values for all units at $A=a$; (3) average
these predicted values over all units.

All of the above estimators involve estimating two nuisance functions
about the conditional means/distributions of the treatment, mediator,
outcome, or posttreatment confounders. Specifically, the g-computation
procedure requires correctly specified models for $\mathbb{E}[Y|x,a,z,m]$
and $P(z|x,a)$; the pure imputation estimator requires correctly
specified models for $\mathbb{E}[Y|x,a,z,m]$ and $\mathbb{E}_{Z|x,a}\mathbb{E}[Y|X,A,Z,M=m]$;
the imputation-then-weighting estimator requires correctly specified
models for $\mathbb{E}[Y|x,a,z,m]$ and $\Pr[A=a|x]$; the pure weighting
estimator requires correctly specified models for $\Pr[A=a|x]$ and
$\Pr[M=m|x,a,z]$; and the weighting-then-imputation estimator requires
correctly specified models for $\Pr[M=m|x,a,z]$ and $\mathbb{E}\big[\frac{\mathbb{I}(M=m)Y}{\Pr[M=m|X,A,Z]}|x,a\big]$.
When either of the two requisite models is misspecified, the resulting
estimator will be inconsistent. Thus, in empirical applications where
the confounders $X$ and $Z$ have many components, these estimators
can be highly prone to model misspecification bias. In the following
section, we introduce four ``doubly robust'' estimators, each of
which requires correct specification of one particular nuisance function
and \textit{either} of two other nuisance functions.

\section{Doubly Robust Estimators\label{sec:Doubly-Robust-Estimators}}

Before proceeding, we introduce the following functions (treating
$a$ and $m$ as fixed):
\begin{align*}
\mu_{y}(x,z) & :=\mathbb{E}[Y|x,a,z,m]\\
\nu_{y}(x) & :=\mathbb{E}_{Z|x,a}\mu_{y}(X,Z)\\
\pi_{a}(x) & :=\Pr[A=a|x]\\
\pi_{m}(x,z) & :=\Pr[M=m|x,a,z],
\end{align*}
Under assumptions 1-3, $\mu_{y}(x,z)=\mathbb{E}[Y(a,m)|x,a,z]$ and
$\nu_{y}(x)=\mathbb{E}[Y(a,m)|x]$. Thus $\mu_{y}(x,z)$ reflects
how the potential outcome $Y(a,m)$ depends on pretreatment confounders
$X$ and posttreatment confounders $Z$ among units with treatment
status $a$, and $\nu_{y}(x)$ reflects how the potential outcome
$Y(a,m)$ depends on pretreatment confounders $X$. Let $\mu_{y}^{\textup{w}}(x,z)$,
$\nu_{y}^{\textup{w}}(x)$, $\pi_{a}^{\textup{w}}(x)$, and $\pi_{m}^{\textup{w}}(x,z$)
denote a set of working models for these nuisance functions, and let
$\hat{\mu}_{y}^{\textup{w}}(x,z)$, $\hat{\nu}_{y}^{\textup{w}}(x)$,
$\hat{\pi}_{a}^{\textup{w}}(x)$, and $\hat{\pi}_{m}^{\textup{w}}(x,z$)
denote their estimates. In particular, consider three different two-step
estimators of $\nu_{y}(x)$:
\begin{align}
\hat{\nu}_{y}^{\textup{w}}(x;\hat{\mu}_{y}^{\textup{w}}) & =\hat{\mathbb{E}}\big[\hat{\mu}_{y}^{\textup{w}}(X,Z)|x,a\big]\label{eq:nu1}\\
\hat{\nu}_{y}^{\textup{w}}(x;\hat{\pi}_{m}^{\textup{w}}) & =\hat{\mathbb{E}}\big[\frac{\mathbb{I}(M=m)Y}{\hat{\pi}_{m}^{\textup{w}}(X,Z)}|x,a\big]\label{eq:nu2}\\
\hat{\nu}_{y}^{\textup{w}}(x;\hat{\mu}_{y}^{\textup{w}},\hat{\pi}_{m}^{\textup{w}}) & =\hat{\mathbb{E}}\big[\hat{\mu}_{y}^{\textup{w}}(X,Z)+\frac{\mathbb{I}(M=m)}{\hat{\pi}_{m}^{\textup{w}}(X,Z)}(Y-\hat{\mu}_{y}^{\textup{w}}(X,Z))|x,a\big],\label{eq:nu3}
\end{align}
where $\hat{\mathbb{E}}[U|x,a]$ denotes estimates of the conditional
mean of a random variable $U$ given $X=x$ and $A=a$. In the above
equations, the notation $\hat{\nu}_{y}^{\textup{w}}(x;\hat{\mu}_{y}^{\textup{w}})$
indicates that this quantity depends on previous estimates of $\mu_{y}^{\textup{w}}(x,z)$,
and the same applies to $\hat{\nu}_{y}^{\textup{w}}(x;\hat{\pi}_{m}^{\textup{w}})$
and $\hat{\nu}_{y}^{\textup{w}}(x;\hat{\mu}_{y}^{\textup{w}},\hat{\pi}_{m}^{\textup{w}})$.
The last expression $\hat{\nu}_{y}^{\textup{w}}(x;\hat{\mu}_{y}^{\textup{w}},\hat{\pi}_{m}^{\textup{w}})$
can be seen as a doubly robust estimator of $\nu_{y}(x)$: when $\nu_{y}(x)$
is correctly specified and either $\mu_{y}^{\textup{w}}(x,z)$ or
$\pi_{m}^{\textup{w}}(x,z)$ is correctly specified, $\hat{\nu}_{y}^{\textup{w}}(x;\hat{\mu}_{y}^{\textup{w}},\hat{\pi}_{m}^{\textup{w}})$
will be consistent for $\nu_{y}(x)$.

Now consider the following estimators of $\psi_{am}$:
\begin{align*}
\hat{\psi}_{am}^{\textup{dr}_{1}} & =\mathbb{P}_{n}\big[\hat{\nu}_{y}^{\textup{w}}(X;\hat{\mu}_{y}^{\textup{w}})+\frac{\mathbb{I}(A=a)}{\hat{\pi}_{a}^{\textup{w}}(X)}\big(\hat{\mu}_{y}^{\textup{w}}(X,Z)-\hat{\nu}_{y}^{\textup{w}}(X;\hat{\mu}_{y}^{\textup{w}})\big)\big]\\
\hat{\psi}_{am}^{\textup{dr}_{2}} & =\mathbb{P}_{n}\big[\hat{\nu}_{y}^{\textup{w}}(X;\hat{\pi}_{m}^{\textup{w}})+\frac{\mathbb{I}(A=a)}{\hat{\pi}_{a}^{\textup{w}}(X)}\big(\frac{\mathbb{I}(M=m)Y}{\hat{\pi}_{m}^{\textup{w}}(X,Z)}-\hat{\nu}_{y}^{\textup{w}}(X;\hat{\pi}_{m}^{\textup{w}})\big)\big]\\
\hat{\psi}_{am}^{\textup{dr}_{3}} & =\mathbb{P}_{n}\Big[\frac{\mathbb{I}(A=a)}{\hat{\pi}_{a}^{\textup{w}}(X)}\Big(\hat{\mu}_{y}^{\textup{w}}(X,Z)+\frac{\mathbb{I}(M=m)}{\hat{\pi}_{m}^{\textup{w}}(X,Z)}\big(Y-\hat{\mu}_{y}^{\textup{w}}(X,Z)\big)\Big)\Big]\\
\hat{\psi}_{am}^{\textup{dr}_{4}} & =\mathbb{P}_{n}\Big[\hat{\nu}_{y}^{\textup{w}}(X;\hat{\mu}_{y}^{\textup{w}},\hat{\pi}_{m}^{\textup{w}})\Big],
\end{align*}
where $\mathbb{P}_{n}[\cdot]=n^{-1}\sum_{i}[\cdot]_{i}.$ $\hat{\psi}_{am}^{\textup{dr}_{1}}$
can be seen as a combination of the pure imputation estimator and
the imputation-then-weighting estimator; $\hat{\psi}_{am}^{\textup{dr}_{2}}$
a combination of the pure weighting estimator and the weighting-then-imputation
estimator; $\hat{\psi}_{am}^{\textup{dr}_{3}}$ a combination of the
pure weighting estimator and the imputation-then-weighting estimator;
and $\hat{\psi}_{am}^{\textup{dr}_{4}}$ a combination of the pure
imputation estimator and the weighting-then-imputation estimator.
Their double robustness is given in Proposition 1.
\begin{prop}
Under assumptions 1-3 and suitable regularity conditions,
\begin{enumerate}
\item $\hat{\psi}_{am}^{\textup{dr}_{1}}$ is CAN if $\mu_{y}^{\textup{w}}(x,z)$
is correctly specified and either $\nu_{y}^{\textup{w}}(x)$ or $\pi_{a}^{\textup{w}}(x)$
is correctly specified.
\item $\hat{\psi}_{am}^{\textup{dr}_{2}}$ is CAN if $\pi_{m}^{\textup{w}}(x,z)$
is correctly specified and either $\nu_{y}^{\textup{w}}(x)$ or $\pi_{a}^{\textup{w}}(x)$
is correctly specified.
\item $\hat{\psi}_{am}^{\textup{dr}_{3}}$ is CAN if $\pi_{a}^{\textup{w}}(x)$
is correctly specified and either $\pi_{m}^{\textup{w}}(x,z)$ or
$\mu_{y}^{\textup{w}}(x,z)$ is correctly specified.
\item $\hat{\psi}_{am}^{\textup{dr}_{4}}$ is CAN if $\nu_{y}^{\textup{w}}(x)$
is correctly specified and either $\pi_{m}^{\textup{w}}(x,z)$ or
$\mu_{y}^{\textup{w}}(x,z)$ is correctly specified.
\end{enumerate}
\end{prop}
The double robustness of these estimators is due to a similar logic
to that of standard doubly robust estimators for the mean of incomplete
data (\citealt{scharfstein1999adjusting,robins2007comment}). For
example, for $\hat{\psi}_{am}^{\textup{dr}_{1}}$, when $\mu_{y}^{\textup{w}}(x,z)$
and $\nu_{y}^{\textup{w}}(x)$ are correctly specified, the second
term inside $\mathbb{P}_{n}[\cdot]$ will have a zero mean (asymptotically),
leaving only $\mathbb{P}_{n}[\hat{\nu}_{y}^{\textup{w}}(X;\hat{\mu}_{y}^{\textup{w}})]$,
i.e., the pure imputation estimator; and when $\mu_{y}^{\textup{w}}(x,z)$
and $\pi_{a}^{\textup{w}}(x)$ are correctly specified, the terms
involving $\hat{\nu}_{y}^{\textup{w}}(X;\hat{\mu}_{y}^{\textup{w}})$
will have a zero mean, leaving only $\mathbb{P}_{n}[\big(\mathbb{I}(A=a)/\hat{\pi}_{a}^{\textup{w}}(X)\big)\hat{\mu}_{y}^{\textup{w}}(X,Z)]$,
i.e., the imputation-then-weighting estimator.

Among these doubly robust estimators, $\hat{\psi}_{am}^{\textup{dr}_{3}}$
can be particularly useful in randomized trials where the treatment
assignment mechanism is known. In this case, $\hat{\psi}_{am}^{\textup{dr}_{3}}$
is consistent as long as either $\pi_{m}^{\textup{w}}(X,Z)$ or $\mu_{y}^{\textup{w}}(X,Z)$
is correctly specified. In observational studies, however, none of
these nuisance functions is known a priori, and the relative utility
of these estimators will depend on the subject matter knowledge the
investigator might have about the data generating process. For example,
if the investigator has a better understanding of the mediator model
than of the outcome models, $\hat{\psi}_{am}^{\textup{dr}_{2}}$ may
be preferred. Yet, in many applications, little information is available
about any part of the data generating process. In those cases, the
multiply robust estimators presented below will be more useful as
they do not hinge on correct specification of any particular nuisance
function. Moreover, as we will see, they are more amenable to the
use of flexible machine learning methods for estimating the nuisance
functions.

\section{Multiply Robust and Semiparametric Efficient Estimators\label{sec:Multiply-Robust}}

\sloppy Henceforth, let $O=(X,A,Z,M,Y)$ denote the observed data,
and $\mathcal{P}_{\textup{np}}$ a nonparametric model over $O$ wherein
all laws satisfy the positivity assumption described in Section \ref{subsec:Notation-Assumptions}.
Define the following of submodels of $\mathcal{P}_{\textup{np}}$:
\begin{itemize}
\item $\mathcal{P}_{1}=\{P\in\mathcal{P}_{\textup{np}}$: $\mu_{y}^{\textup{w}}(x,z)$
and $\pi_{a}^{\textup{w}}(x)$ are correctly specified$\}$
\item $\mathcal{P}_{2}=\{P\in\mathcal{P}_{\textup{np}}$: $\mu_{y}^{\textup{w}}(x,z)$
and $\nu_{y}^{\textup{w}}(x)$ are correctly specified$\}$
\item $\mathcal{P}_{3}=\{P\in\mathcal{P}_{\textup{np}}$: $\pi_{m}^{\textup{w}}(x,z)$
and $\pi_{a}^{\textup{w}}(x)$ are correctly specified$\}$
\item $\mathcal{P}_{4}=\{P\in\mathcal{P}_{\textup{np}}$: $\pi_{m}^{\textup{w}}(x,z)$
and $\nu_{y}^{\textup{w}}(x)$ are correctly specified$\}$
\end{itemize}
Consider the following estimators of $\psi_{am}$:
\begin{align*}
\hat{\psi}_{am}^{\textup{tr}_{1}} & =\mathbb{P}_{n}\Big[\hat{\nu}_{y}^{\textup{w}}(X;\hat{\mu}_{y}^{\textup{w}})+\frac{\mathbb{I}(A=a)}{\hat{\pi}_{a}^{\textup{w}}(X)}\big(\hat{\mu}_{y}^{\textup{w}}(X,Z)-\hat{\nu}_{y}^{\textup{w}}(X;\hat{\mu}_{y}^{\textup{w}})\big)+\frac{\mathbb{I}(A=a)\mathbb{I}(M=m)}{\hat{\pi}_{a}^{\textup{w}}(X)\hat{\pi}_{m}^{\textup{w}}(X,Z)}\big(Y-\hat{\mu}_{y}^{\textup{w}}(X,Z)\big)\Big]\\
\hat{\psi}_{am}^{\textup{tr}_{2}} & =\mathbb{P}_{n}\Big[\hat{\nu}_{y}^{\textup{w}}(X;\hat{\pi}_{m}^{\textup{w}})+\frac{\mathbb{I}(A=a)}{\hat{\pi}_{a}^{\textup{w}}(X)}\big(\hat{\mu}_{y}^{\textup{w}}(X,Z)-\hat{\nu}_{y}^{\textup{w}}(X;\hat{\pi}_{m}^{\textup{w}})\big)+\frac{\mathbb{I}(A=a)\mathbb{I}(M=m)}{\hat{\pi}_{a}^{\textup{w}}(X)\hat{\pi}_{m}^{\textup{w}}(X,Z)}\big(Y-\hat{\mu}_{y}^{\textup{w}}(X,Z)\big)\Big]\\
\hat{\psi}_{am}^{\textup{qr}} & =\mathbb{P}_{n}\Big[\hat{\nu}_{y}^{\textup{w}}(X;\hat{\mu}_{y}^{\textup{w}},\hat{\pi}_{m}^{\textup{w}})+\frac{\mathbb{I}(A=a)}{\hat{\pi}_{a}^{\textup{w}}(X)}\big(\hat{\mu}_{y}^{\textup{w}}(X,Z)-\hat{\nu}_{y}^{\textup{w}}(X;\hat{\mu}_{y}^{\textup{w}},\hat{\pi}_{m}^{\textup{w}})\big)+\frac{\mathbb{I}(A=a)\mathbb{I}(M=m)}{\hat{\pi}_{a}^{\textup{w}}(X)\hat{\pi}_{m}^{\textup{w}}(X,Z)}\big(Y-\hat{\mu}_{y}^{\textup{w}}(X,Z)\big)\Big]
\end{align*}
The triple robustness of $\hat{\psi}_{am}^{\textup{tr}_{1}}$ and
$\hat{\psi}_{am}^{\textup{tr}_{2}}$ and the quadruple robustness
of $\hat{\psi}_{am}^{\textup{qr}}$ are given below.
\begin{prop}
Under assumptions 1-3 and suitable regularity conditions, $\hat{\psi}_{am}^{\textup{tr}_{1}}$
is CAN in $\mathcal{P}_{1}\cup\mathcal{P}_{2}\cup\mathcal{P}_{3}$,
$\hat{\psi}_{am}^{\textup{tr}_{2}}$ is CAN in $\mathcal{P}_{1}\cup\mathcal{P}_{3}\cup\mathcal{P}_{4}$,
and $\hat{\psi}_{am}^{\textup{qr}}$ is CAN in $\mathcal{P}_{1}\cup\mathcal{P}_{2}\cup\mathcal{P}_{3}\cup\mathcal{P}_{4}$.
In addition, $\hat{\psi}_{am}^{\textup{tr}_{1}}$, $\hat{\psi}_{am}^{\textup{tr}_{2}}$,
and $\hat{\psi}_{am}^{\textup{qr}}$ are all locally efficient in
the sense that they attain the semiparametric efficiency bound of
$\mathcal{P}_{\textup{np}}$ at $\mathcal{P}_{1}\cap\mathcal{P}_{3}$,
i.e., when all of the four nuisance functions are correctly specified.
\end{prop}
The multiple robustness of these estimators is due to a similar logic
to that of the doubly robust estimators given previously. For example,
for $\hat{\psi}_{am}^{\textup{qr}}$, when $\mu_{y}^{\textup{w}}(x,z)$
and $\pi_{a}^{\textup{w}}(x)$ are correctly specified ($\mathcal{P}_{1}$),
the terms involving $\hat{\nu}_{y}^{\textup{w}}(X;\hat{\mu}_{y}^{\textup{w}},\hat{\pi}_{m}^{\textup{w}})$
and the third term inside $\mathbb{P}_{n}[\cdot]$ will both have
a zero mean (asymptotically), leaving only $\mathbb{P}_{n}[\big(\mathbb{I}(A=a)/\hat{\pi}_{a}^{\textup{w}}(X)\big)\hat{\mu}_{y}^{\textup{w}}(X,Z)]$,
the imputation-then-weighting estimator; when $\mu_{y}^{\textup{w}}(x,z)$
and $\nu_{y}^{\textup{w}}(x)$ are correctly specified ($\mathcal{P}_{2}$),
both the second and third terms inside $\mathbb{P}_{n}[\cdot]$ will
have a zero mean, leaving only $\mathbb{P}_{n}[\hat{\nu}_{y}^{\textup{w}}(X;\hat{\mu}_{y}^{\textup{w}},\hat{\pi}_{m}^{\textup{w}})]$,
i.e., the doubly robust estimator $\hat{\psi}_{am}^{\textup{dr}_{4}}$
; when $\pi_{m}^{\textup{w}}(x,z)$ and $\pi_{a}^{\textup{w}}(x)$
are correctly specified ($\mathcal{P}_{3}$), the terms involving
$\hat{\mu}_{y}^{\textup{w}}(X,Z)$ and $\hat{\nu}_{y}^{\textup{w}}(X;\hat{\mu}_{y}^{\textup{w}},\hat{\pi}_{m}^{\textup{w}})$
will both have a zero mean, leaving only $\mathbb{P}_{n}\big[\big(\mathbb{I}(A=a)\mathbb{I}(M=m)\big)Y/\big(\hat{\pi}_{a}^{\textup{w}}(X)\hat{\pi}_{m}^{\textup{w}}(X,Z)\big)\big]$,
i.e., the pure weighting estimator; and when $\pi_{m}^{\textup{w}}(x,z)$
and $\nu_{y}^{\textup{w}}(x)$ are correctly specified ($\mathcal{P}_{4}$),
the terms involving $\hat{\mu}_{y}^{\textup{w}}(X,Z)$ and $\hat{\pi}_{a}^{\textup{w}}(X)$
will both have a zero mean, leaving only $\mathbb{P}_{n}[\hat{\nu}_{y}^{\textup{w}}(X;\hat{\mu}_{y}^{\textup{w}},\hat{\pi}_{m}^{\textup{w}})]$,
i.e., the doubly robust estimator $\hat{\psi}_{am}^{\textup{dr}_{4}}$.

The asymptotic efficiency of these estimators is due to the fact that
they all solve the estimating equation formed by the efficient influence
function of $\psi_{am}$, which is 
\begin{equation}
\varphi_{am}(O)=\nu_{y}(X)+\frac{\mathbb{I}(A=a)}{\pi_{a}(X)}\big(\mu_{y}(X,Z)-\nu_{y}(X)\big)+\frac{\mathbb{I}(A=a)\mathbb{I}(M=m)}{\pi_{a}(X)\pi_{m}(X,Z)}\big(Y-\mu_{y}(X,Z)\big)-\psi_{am},\label{eq:EIF}
\end{equation}
and the fact that $\mathbb{E}[\varphi_{am}(O;\eta)]$ has a zero derivative
with respect to the nuisance functions $\eta=(\mu_{y}^{\textup{w}}(x,z),\nu_{y}^{\textup{w}}(x),\pi_{a}^{\textup{w}}(x),\pi_{m}^{\textup{w}}(x,z))$
at the truth (for a derivation of this influence function in the context
of time-varying treatments, see \citealt{van2012targeted}). The latter
property implies that first step estimation of the nuisance functions
will have no (first-order) effect on the influence function of $\hat{\psi}_{am}^{\textup{tr}_{1}}$,
$\hat{\psi}_{am}^{\textup{tr}_{2}}$, and $\hat{\psi}_{am}^{\textup{qr}}$.
In practice, the nuisance functions can be estimated via data-adaptive
methods combined with cross-fitting (\citealt{zheng2011cross,chernozhukov2018double}),
in which case estimates of $\psi_{am}$ (and hence CDE) are semiparametric
efficient when estimates of the nuisance functions, for example, all
converge at faster-than-$n^{-1/4}$ rates.\footnote{More precisely, $\hat{\psi}_{am}^{\textup{tr}_{1}}$, $\hat{\psi}_{am}^{\textup{tr}_{2}}$,
and $\hat{\psi}_{am}^{\textup{qr}}$ are semiparametric efficient
if $R_{n}(\hat{\pi}_{a})R_{n}(\hat{\nu}_{y})+R_{n}(\hat{\pi}_{m})R_{n}(\hat{\mu}_{y})=o(n^{-1/2})$,
where $R_{n}(\cdot)$ maps a nuisance function to its $L_{2}(P)$
convergence rate with respect to the true distribution $P$. See Supporting
Material C or \citet{rotnitzky2017multiply}.}

Among the above estimators, $\hat{\psi}_{am}^{\textup{tr}_{1}}$ is
akin to the estimator proposed by \citet{bang2005doubly} for the
mean of a potential outcome with time-varying treatments and time-varying
confounders. Specifically, they suggest that $\mathbb{I}(A=a)\mathbb{I}(M=m)Y/\big(\hat{\pi}_{a}^{\textup{w}}(X)\hat{\pi}_{m}^{\textup{w}}(X,Z)\big)$
be included as a covariate in a generalized linear model (with canonical
link) for $\hat{\mu}_{y}^{\textup{w}}(x,z)$, and $\mathbb{I}(A=a)/\hat{\pi}_{a}^{\textup{w}}(X)$
be included as a covariate in a generalized linear model (with canonical
link) for $\hat{\nu}_{y}^{\textup{w}}(x;\hat{\mu}_{y}^{\textup{w}})$,
in which case the score equations ensure that both the second and
third terms inside $\mathbb{P}_{n}[\cdot]$ have a \textit{zero sample
mean, }thus leaving only $\mathbb{P}_{n}[\hat{\nu}_{y}^{\textup{w}}(X;\hat{\mu}_{y}^{\textup{w}})]$,
i.e., the pure imputation estimator. Because this procedure estimates
$\psi_{am}$ as a sample average of $\hat{\nu}_{y}^{\textup{w}}(X;\hat{\mu}_{y}^{\textup{w}})$,
which typically resides in the parameter space of $\psi_{am}$, it
tends to be more stable in finite samples than the unadjusted estimator
$\hat{\psi}_{am}^{\textup{tr}_{1}}$ (\citealt{robins2007comment}).
$\hat{\psi}_{am}^{\textup{tr}_{2}}$ and $\hat{\psi}_{am}^{\textup{qr}}$
differ from $\hat{\psi}_{am}^{\textup{tr}_{1}}$ only in the response
variable they use to model $\nu_{y}(X)$: $\hat{\psi}_{am}^{\textup{tr}_{2}}$
uses $\mathbb{I}(M=m)/\hat{\pi}_{m}^{\textup{w}}(X,Z)$ whereas $\hat{\psi}_{am}^{\textup{qr}}$
uses $\hat{\mu}_{y}^{\textup{w}}(X,Z)+\mathbb{I}(M=m)(Y-\hat{\mu}_{y}^{\textup{w}}(X,Z))/\hat{\pi}_{m}^{\textup{w}}(X,Z)$,
which adds another layer of robustness. In fact, $\hat{\psi}_{am}^{\textup{qr}}$
constitutes a special case of the $2^{K}$-robust estimator proposed
by \citet{rotnitzky2017multiply} in the context of time-varying treatments.
In practice, the Bang-Robins procedure can also be applied to $\hat{\psi}_{am}^{\textup{tr}_{2}}$
and $\hat{\psi}_{am}^{\textup{qr}}$ to improve their finite-sample
performance.

When flexible machine learning methods (instead of generalized linear
models) are used to estimate the nuisance functions, the Bang-Robins
procedure can no longer ensure a zero sample mean of the second and
third terms inside $\mathbb{P}_{n}[\cdot]$. In this case, the method
of targeted maximum likelihood estimation (TMLE; \citealt{van2006targeted})
can be used to adjust the first step estimates of $\mu_{y}(x,z)$
and $\nu_{y}(x)$ such that the second and third terms inside $\mathbb{P}_{n}[\cdot]$
have a zero sample mean. This approach may yield better finite-sample
performance than the unadjusted estimators and more robustness than
the Bang-Robins procedure based on generalized linear models.

For inference of $\hat{\psi}_{am}^{\textup{tr}_{1}}$, $\hat{\psi}_{am}^{\textup{tr}_{2}}$,
$\hat{\psi}_{am}^{\textup{qr}}$, and the corresponding estimates
of the CDE, the nonparametric bootstrap can be used when the nuisance
functions are estimated using parametric models. When data-adaptive
methods are used to estimate the nuisance functions, it will be reasonable
to use the empirical analog of the efficient influence function to
construct standard errors and Wald-type confidence intervals. For
example, the variance of $\widehat{\textup{CDE}}(a,a',m)$ can be
estimated by $\mathbb{P}_{n}[(\hat{\varphi}_{am}(O)-\hat{\varphi}_{a'm}(O))^{2}]/n$.
When the CDE is defined on the risk ratio or odds ratio scale, corresponding
variance estimates can be obtained using the delta method.
\begin{figure}[!th]
\centering{}\includegraphics[width=1\textwidth]{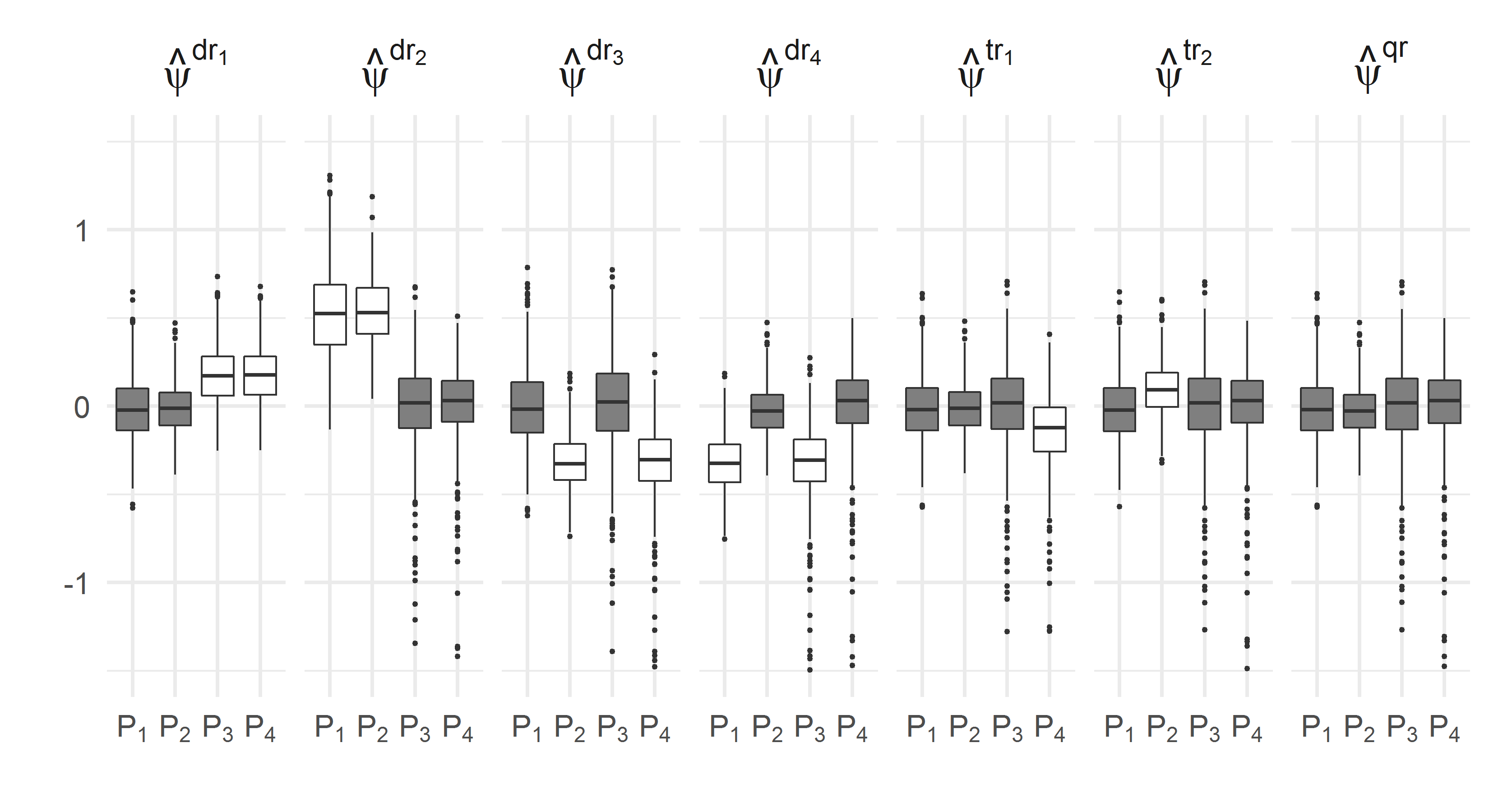}\caption{Sampling distributions of the doubly and multiply robust estimators
for $n=2,000$.\label{fig:simulation}}
\end{figure}

\section{Monte Carlo Evidence}

We now present a simulation study to demonstrate the multiple robustness
of the proposed estimators. The data generating process is similar
to that used in \citet{miles2020semiparametric} and is described
in greater detail in Supporting Material D. We generate 1,000 Monte
Carlo samples of size 2,000, and, without loss of generality, focus
on the estimand $\psi_{01}=\mathbb{E}[Y(0,1)]$. We examine the sampling
distributions of all of the doubly and multiply robust estimators
described above under conditions associated with submodels $\mathcal{P}_{1}$,
$\mathcal{P}_{2}$, $\mathcal{P}_{3}$, $\mathcal{P}_{4}$. The results
are shown in Figure \ref{fig:simulation}, where each panel corresponds
to an estimator, and the y axis is recentered at the true value of
$\psi_{01}$. The shaded box plots highlight the cases under which
a given estimator should be consistent. We can see that all of the
doubly and multiply robust estimators behave as expected. They center
around the true value if and only if the requisite nuisance functions
are all correctly specified.

\bibliographystyle{elsarticle-harv}
\bibliography{causality_ref}

\clearpage{}

\appendix

\section{Proof of Equations (\ref{eq:g-comp}-\ref{eq:weighting}).}

Starting from equation \eqref{eq:weighting-then-imputation}, we have
\begin{align}
 & \mathbb{E}_{X}\mathbb{E}\big[\frac{\mathbb{I}(M=m)Y}{\Pr[M=m|X,A,Z]}|X,A=a\big]\nonumber \\
= & \mathbb{E}_{X}\Big[\mathbb{E}\big[\frac{\mathbb{I}(A=a)\mathbb{I}(M=m)Y}{\Pr[A=a|X]\Pr[M=m|X,A,Z]}|X,A=a\big]\cdot\Pr[A=a|X]+0\cdot\Pr[A=a|X]\Big]\nonumber \\
= & \mathbb{E}[\frac{\mathbb{I}(A=a)\mathbb{I}(M=m)Y}{\Pr[A=a|X]\Pr[M=m|X,A,Z]}]\label{eq:weighting2}\\
= & \mathbb{E}\big[\frac{\mathbb{I}(A=a)}{\Pr[A=a|X]}\mathbb{E}[\frac{\mathbb{I}(M=m)Y}{\Pr[M=m|X,A,Z]}|X,A,Z]\big]\nonumber \\
= & \mathbb{E}\big[\frac{\mathbb{I}(A=a)}{\Pr[A=a|X]}\mathbb{E}[\frac{\mathbb{I}(M=m)Y}{\Pr[M=m|X,A,Z]}|X,A,Z,M=m]\Pr[M=m|X,A=a,Z]\big]\nonumber \\
= & \mathbb{E}\big[\frac{\mathbb{I}(A=a)\mathbb{E}[Y|X,A,Z,M=m]}{\Pr[A=a|X]}\big]\label{eq:imputation-then-weighting2}\\
= & \mathbb{E}_{X,A}\mathbb{E}\big[\frac{\mathbb{I}(A=a)\mathbb{E}[Y|X,A,Z,M=m]}{\Pr[A=a|X]}|X,A\big]\nonumber \\
= & \mathbb{E}_{X,A}\Big[\mathbb{E}\big[\frac{\mathbb{I}(A=a)\mathbb{E}[Y|X,A,Z,M=m]}{\Pr[A=a|X]}|X,A=a\big]\Pr[A=a|X]\Big]\nonumber \\
= & \mathbb{E}_{X}\mathbb{E}\big[\mathbb{E}[Y|X,A,Z,M=m]|X,A=a\big]\label{eq:imputation2}\\
= & \iiint\mathbb{E}[Y|x,a,z,m]dP(z|x,a)dP(x).\label{eq:g-comp2}
\end{align}
Equations \eqref{eq:weighting2}, \eqref{eq:imputation-then-weighting2},
\eqref{eq:imputation2}, and \eqref{eq:g-comp2} correspond to equations
\eqref{eq:weighting}, \eqref{eq:imputation-weighting}, \eqref{eq:imputation},
and \eqref{eq:g-comp}, respectively.

\section{Proof of Proposition 1}

Below we show that $\hat{\psi}_{am}^{\textup{dr}_{1}}$ is CAN when
(a) $\mu_{y}^{\textup{w}}(x,z)$ is correctly specified and (b) either
$\nu_{y}^{\textup{w}}(x)$ or $\pi_{a}^{\textup{w}}(x)$ is correctly
specified. The double robustness of $\hat{\psi}_{am}^{\textup{dr}_{2}}$,
$\hat{\psi}_{am}^{\textup{dr}_{3}}$, and $\hat{\psi}_{am}^{\textup{dr}_{4}}$
can be verified analogously.

A first-order Taylor expansion of $\hat{\psi}_{am}^{\textup{dr}_{1}}$
implies that
\[
\hat{\psi}_{am}^{\textup{dr}_{1}}=\mathbb{P}_{n}\big[\nu_{y}^{*}(X)+\frac{\mathbb{I}(A=a)}{\pi_{a}^{*}(X)}\big(\mu_{y}^{*}(X,Z)-\nu_{y}^{*}(X)\big)\big]+o_{p}(1),
\]
where $\nu_{y}^{*}(X)$, $\pi_{a}^{*}(X)$ and $\mu_{y}^{*}(X,Z)$
denote the probability limits of $\hat{\nu}_{y}^{\textup{w}}(X)$,
$\hat{\pi}_{a}^{\textup{w}}(X)$, and $\hat{\mu}_{y}^{\textup{w}}(X,Z)$.
Hence it suffices to show $\mathbb{E}\big[\nu_{y}^{*}(X)+\frac{\mathbb{I}(A=a)}{\pi_{a}^{*}(X)}\big(\mu_{y}^{*}(X,Z)-\nu_{y}^{*}(X)\big)\big]=\psi_{am}$
if $\mu_{y}^{*}(X,Z)=\mu_{y}(X,Z)$ and either $\pi_{a}^{*}(X)=\pi_{a}(X)$
or $\nu_{y}^{*}(X)=\nu_{y}(X)$. Consistency follows from the law
of large numbers, and asymptotic normality follows from standard regularity
conditions for M-estimators.

When $\mu_{y}^{*}(X,Z)=\mu_{y}(X,Z)$ and $\pi_{a}^{*}(X)=\pi_{a}(X)$,
\begin{align*}
\textup{plim}\,\hat{\psi}_{am}^{\textup{dr}_{1}}= & \mathbb{E}\big[\nu_{y}^{*}(X)+\frac{\mathbb{I}(A=a)}{\pi_{a}(X)}\big(\mu_{y}(X,Z)-\nu_{y}^{*}(X)\big)\big]\\
= & \mathbb{E}_{X}\mathbb{E}\big[\nu_{y}^{*}(X)+\frac{\mathbb{I}(A=a)}{\pi_{a}(X)}\big(\mu_{y}(X,Z)-\nu_{y}^{*}(X)\big)|X\big]\\
= & \mathbb{E}_{X}\big[\nu_{y}^{*}(X)+\mathbb{E}[\mu_{y}(X,Z)-\nu_{y}^{*}(X)|X,A=a]\big]\\
= & \mathbb{E}_{X}\big[\nu_{y}^{*}(X)+\mathbb{E}[\mu_{y}(X,Z)|X,A=a]-\nu_{y}^{*}(X)\big]\\
= & \mathbb{E}_{X}\mathbb{E}[\mu_{y}(X,Z)|X,A=a]\\
= & \psi_{am}.
\end{align*}
When $\mu_{y}^{*}(X,Z)=\mu_{y}(X,Z)$ and $\nu_{y}^{*}(X)=\nu_{y}(X)$,
\begin{align*}
\textup{plim}\,\hat{\psi}_{am}^{\textup{dr}_{1}}= & \mathbb{E}\big[\nu_{y}(X)+\frac{\mathbb{I}(A=a)}{\pi_{a}^{*}(X)}\big(\mu_{y}(X,Z)-\nu_{y}(X)\big)\big]\\
= & \mathbb{E}_{X}\mathbb{E}\big[\nu_{y}(X)+\frac{\mathbb{I}(A=a)}{\pi_{a}^{*}(X)}\big(\mu_{y}(X,Z)-\nu_{y}(X)\big)|X\big]\\
= & \mathbb{E}_{X}\big[\nu_{y}(X)+\mathbb{E}[\frac{\mathbb{I}(A=a)}{\pi_{a}^{*}(X)}\big(\mu_{y}(X,Z)-\nu_{y}(X)\big)|X,A=a]\pi_{a}(X)\big]\\
= & \mathbb{E}_{X}\big[\nu_{y}(X)+\frac{\pi_{a}(X)}{\pi_{a}^{*}(X)}\mathbb{E}[\mu_{y}(X,Z)-\nu_{y}(X)|X,A=a]\big]\\
= & \mathbb{E}_{X}\big[\nu_{y}(X)+0\big]\\
= & \psi_{am}.
\end{align*}

\section{Proof of Proposition 2}

Below we show that $\hat{\psi}_{am}^{\textup{tr}_{1}}$ is CAN in
$\mathcal{P}_{1}\cup\mathcal{P}_{2}\cup\mathcal{P}_{3}$ and locally
efficient in $\mathcal{P}_{1}\cap\mathcal{P}_{3}$. The multiple robustness
and local efficiency of $\hat{\psi}_{am}^{\textup{tr}_{2}}$ and $\hat{\psi}_{am}^{\textup{qr}}$
can be verified analogously.

A first-order Taylor expansion of $\hat{\psi}_{am}^{\textup{tr}_{1}}$
implies that
\[
\hat{\psi}_{am}^{\textup{tr}_{1}}=\mathbb{P}_{n}\big[\nu_{y}^{*}(X)+\frac{\mathbb{I}(A=a)}{\pi_{a}^{*}(X)}\big(\mu_{y}^{*}(X,Z)-\nu_{y}^{*}(X)\big)+\frac{\mathbb{I}(A=a)\mathbb{I}(M=m)}{\pi_{a}^{*}(X)\pi_{m}^{*}(X,Z)}\big(Y-\mu_{y}^{*}(X,Z)\big)\big]+o_{p}(1),
\]
where $\mu_{y}^{*}(X,Z)$, $\nu_{y}^{*}(X)$, $\pi_{a}^{*}(X)$, and
$\pi_{m}^{*}(X,Z)$ denote the probability limits of $\hat{\mu}_{y}^{\textup{w}}(X,Z)$,
$\hat{\nu}_{y}^{\textup{w}}(X)$, $\hat{\pi}_{a}^{\textup{w}}(X)$,
and $\hat{\pi}_{m}(X,Z)$. Hence it suffices to show that the expectation
of the quantity inside $\mathbb{P}_{n}[\cdot]$ equals $\psi_{am}$
in $\mathcal{P}_{1}\cup\mathcal{P}_{2}\cup\mathcal{P}_{3}$. Consistency
follows from the law of large numbers, and asymptotic normality follows
from standard regularity conditions for M-estimators. First, consider
submodel $\mathcal{P}_{2}$, under which we have $\mu_{y}^{*}(X,Z)=\mu_{y}(X,Z)$
and $\nu_{y}^{*}(X)=\nu_{y}(X)$. From the proof of proposition 1,
we know that 
\[
\mathbb{E}\Big[\nu_{y}(X)+\frac{\mathbb{I}(A=a)}{\pi_{a}^{*}(X)}\big(\mu_{y}(X,Z)-\nu_{y}(X)\big)\Big]=\psi_{am}.
\]
Thus
\begin{align*}
\textup{plim}\,\hat{\psi}_{am}^{\textup{tr}_{1}} & =\psi_{am}+\mathbb{E}\big[\frac{\mathbb{I}(A=a)\mathbb{I}(M=m)}{\pi_{a}^{*}(X)\pi_{m}^{*}(X,Z)}\big(Y-\mu_{y}(X,Z)\big)\big]\\
 & =\psi_{am}+\mathbb{E}_{X}\mathbb{E}\big[\frac{\mathbb{I}(A=a)\mathbb{I}(M=m)}{\pi_{a}^{*}(X)\pi_{m}^{*}(X,Z)}\big(Y-\mu_{y}(X,Z)\big)\big\vert X\big]\\
 & =\psi_{am}+\mathbb{E}_{X}\Big[\mathbb{E}\big[\frac{\mathbb{I}(A=a)\mathbb{I}(M=m)}{\pi_{a}^{*}(X)\pi_{m}^{*}(X,Z)}\big(Y-\mu_{y}(X,Z)\big)\big)|X,A=a\big]\pi_{a}(X)\Big]\\
 & =\psi_{am}+\mathbb{E}_{X}\Big[\frac{\pi_{a}(X)}{\pi_{a}^{*}(X)}\mathbb{E}\big[\frac{\mathbb{I}(M=m)}{\pi_{m}^{*}(X,Z)}\big(Y-\mu_{y}(X,Z)\big)|X,A=a\big]\Big]\\
 & =\psi_{am}+\mathbb{E}_{X}\Big[\frac{\pi_{a}(X)}{\pi_{a}^{*}(X)}\mathbb{E}_{Z|X,A=a}\mathbb{E}\big[\frac{\mathbb{I}(M=m)\pi_{m}(X,Z)}{\pi_{m}^{*}(X,Z)}\big(Y-\mu_{y}(X,Z)\big)|X,A=a,Z,M=m\big]\Big]\\
 & =\psi_{am}+\mathbb{E}_{X}\Big[\frac{\pi_{a}(X)}{\pi_{a}^{*}(X)}\mathbb{E}_{Z|X,A=a}\frac{\pi_{m}(X,Z)}{\pi_{m}^{*}(X,Z)}\underbrace{\mathbb{E}\big[\big(Y-\mu_{y}(X,Z)\big)|X,A=a,Z,M=m\big]}_{=0}\Big]\\
 & =\psi_{am}.
\end{align*}
Then, under $\mathcal{P}_{1}$, we have $\mu_{y}^{*}(X,Z)=\mu_{y}(X,Z)$
and $\pi_{a}^{*}(X)=\pi_{a}(X)$. From the proof of proposition 1,
we know that
\[
\mathbb{E}\Big[\nu_{y}^{*}(X)+\frac{\mathbb{I}(A=a)}{\pi_{a}(X)}\big(\mu_{y}(X,Z)-\nu_{y}^{*}(X)\big)\big]=\psi_{am}.
\]
Thus $\textup{plim}\,\hat{\psi}_{am}^{\textup{tr}_{1}}=\psi_{am}+\mathbb{E}\big[\frac{\mathbb{I}(A=a)\mathbb{I}(M=m)}{\pi_{a}^{*}(X)\pi_{m}^{*}(X,Z)}\big(Y-\mu_{y}(X,Z)\big)\big]=0$
(directly from the above proof for submodel $\mathcal{P}_{2}$). Finally,
under $\mathcal{P}_{3}$, we have $\pi_{a}^{*}(X)=\pi_{a}(X)$ and
$\pi_{m}^{*}(X,Z)=\pi_{m}(X,Z)$.
\begin{align*}
\textup{plim}\,\hat{\psi}_{am}^{\textup{tr}_{1}} & =\mathbb{E}\big[\nu_{y}^{*}(X)+\frac{\mathbb{I}(A=a)}{\pi_{a}(X)}\big(\mu_{y}^{*}(X,Z)-\nu_{y}^{*}(X)\big)+\frac{\mathbb{I}(A=a)\mathbb{I}(M=m)}{\pi_{a}(X)\pi_{m}(X,Z)}\big(Y-\mu_{y}^{*}(X,Z)\big)\big]\\
 & =\mathbb{E}_{X}[\nu_{y}^{*}(X)]+\mathbb{E}_{X}\Big[\mathbb{E}\big[\frac{\mathbb{I}(A=a)}{\pi_{a}(X)}\big(\mu_{y}^{*}(X,Z)-\nu_{y}^{*}(X)\big)|X,A=a\big]\pi_{a}(X)\Big]\\
 & +\mathbb{E}_{X}\mathbb{E}_{Z|X,A=a}\Big[\mathbb{E}\big[\frac{\mathbb{I}(A=a)\mathbb{I}(M=m)}{\pi_{a}(X)\pi_{m}(X,Z)}\big(Y-\mu_{y}^{*}(X,Z)\big)|X,A=a,Z\big]\pi_{a}(X)\Big]\\
 & =\mathbb{E}_{X}[\nu_{y}^{*}(X)]+\mathbb{E}_{X}\mathbb{E}\big[\big(\mu_{y}^{*}(X,Z)-\nu_{y}^{*}(X)\big)|X,A=a\big]\\
 & +\mathbb{E}_{X}\mathbb{E}_{Z|X,A=a}\Big[\mathbb{E}\big[\frac{\mathbb{I}(A=a)\mathbb{I}(M=m)}{\pi_{a}(X)\pi_{m}(X,Z)}\big(Y-\mu_{y}^{*}(X,Z)\big)|X,A=a,Z,M=m\big]\pi_{m}(X,Z)\pi_{a}(X)\Big]\\
 & =\mathbb{E}_{X}\Big[\mu_{y}^{*}(X,Z)+\mathbb{E}_{Z|X,A=a}\mathbb{E}\big[\big(Y-\mu_{y}^{*}(X,Z)\big)|X,A=a,Z,M=m\big]\Big]\\
 & =\mathbb{E}_{X}\Big[\mu_{y}^{*}(X,Z)+\mu_{y}(X,Z)-\mu_{y}^{*}(X,Z)\Big]\\
 & =\psi_{am}.
\end{align*}

To show that $\hat{\psi}_{am}^{\textup{tr}_{1}}$ is locally efficient,
we first verify that equation \eqref{eq:EIF} is the efficient influence
function of $\psi_{am}$ in $\mathcal{P}_{\textup{np}}$, i.e., 
\begin{equation}
\frac{\partial\psi_{am}(t)}{\partial t}\biggl\vert_{t=0}=\mathbb{E}[\varphi_{a,m}^{\textup{eff}}(O)S_{0}(O)],\label{eq:proof}
\end{equation}
where $S_{0}(O)$ is the score function for any one-dimensional submodel
$P_{t}(O)$ evaluated at $t=0$. We first note that $S_{t}(O)$ can
be written as $S_{t}(O)=S_{t}(X)+S_{t}(A|X)+S_{t}(Z|X,A)+S_{t}(M|X,A,Z)+S_{t}(Y|X,A,Z,M)$,
where $S_{t}(u|v)=\partial\log p_{t}(u|v)/\partial t$ and $p_{t}(u|v)$
is the conditional probability density/mass function of $U$ given
$V$. Using equation \eqref{eq:identification} and the product rule,
the left hand side of equation \eqref{eq:proof} can be written as
\begin{align*}
\frac{\partial\psi_{am}(t)}{\partial t}\biggl\vert_{t=0} & =\frac{\partial\iiint ydP_{t}(y|x,a,z,m)dP_{t}(z|x,a)dP_{t}(x)}{\partial t}\biggl\vert_{t=0}\\
 & =\underbrace{\iiint yS_{0}(x)dP_{0}(y|x,a,z,m)dP_{0}(z|x,a)dP_{0}(x)}_{=:\textup{\ensuremath{\phi_{1}}}}\\
 & +\underbrace{\iiint yS_{0}(z|x,a)dP_{0}(y|x,a,z,m)dP_{0}(z|x,a)dP_{0}(x)}_{=:\phi_{2}}\\
 & +\underbrace{\iiint yS_{0}(y|x,a,z,m)dP_{0}(y|x,a,z,m)dP_{0}(z|x,a)dP_{0}(x)}_{=:\phi_{3}}\\
 & =\phi_{1}+\phi_{2}+\phi_{3}
\end{align*}
where the second equality follows from the fact that $\partial dP_{t}(u|v)/\partial t=S_{t}(u|v)dP_{t}(u|v).$

Before evaluating the right hand side of equation \eqref{eq:proof},
we introduce the following shorthands:
\begin{align*}
\varphi_{1}(X) & =\nu_{y}(X),\\
\varphi_{2}(X,A,Z) & =\frac{\mathbb{I}(A=a)}{\pi_{a}(X)}\big(\mu_{y}(X,Z)-\nu_{y}(X)\big),\\
\varphi_{3}(X,A,Z,M,Y) & =\frac{\mathbb{I}(A=a)\mathbb{I}(M=m)}{\pi_{a}(X)\pi_{m}(X,Z)}\big(Y-\mu_{y}(X,Z)\big).
\end{align*}
Thus $\varphi_{a,m}^{\textup{eff}}(O)=\varphi_{1}(X)+\varphi_{2}(X,A,Z)+\varphi_{3}(X,A,Z,M,Y)-\psi_{am}$.
We first observe
\begin{align}
 & \mathbb{E}[\varphi_{1}(X)S_{0}(O)]\nonumber \\
= & \mathbb{E}[\varphi_{1}(X)\big(S_{0}(X)+S_{0}(A|X)+S_{0}(Z|X,A)+S_{0}(M|X,A,Z)+S_{0}(Y|X,A,Z,M)\big)]\nonumber \\
= & \mathbb{E}[\varphi_{1}(X)S_{0}(X)]+\mathbb{E}[\varphi_{1}(X)S_{0}(A|X)]\ldots+\mathbb{E}[\varphi_{1}(X)S_{0}(Y|X,A,Z,M)]\nonumber \\
= & \mathbb{E}[\varphi_{1}(X)S_{0}(X)]+\mathbb{E}\big[\varphi_{1}(X)\underbrace{\mathbb{E}[S_{0}(A|X)|X]}_{=0}\big]\ldots+\mathbb{E}\big[\varphi_{1}(X)\underbrace{\mathbb{E}[S_{0}(Y|X,A,Z,M)|X,A,Z,M]}_{=0}\big]\nonumber \\
= & \mathbb{E}[\varphi_{1}(X)S_{0}(X)]\nonumber \\
= & \mathbb{E}[\mathbb{E}_{Z|X,A=a}\mathbb{E}[Y|X,A=a,Z,M=m]S_{0}(X)]\nonumber \\
= & \phi_{1}\label{eq:psi1}
\end{align}
where we used the fact that $\mathbb{E}[S(U|V)|V]=0$ for any score
function $S(U,V)$. Second,
\begin{align}
 & \mathbb{E}[\varphi_{2}(X,A,Z)S_{0}(O)]\nonumber \\
= & \mathbb{E}[\varphi_{2}(X,A,Z)\big(S_{0}(X)+S_{0}(A|X)+S_{0}(Z|X,A)+S_{0}(M|X,A,Z)+S_{0}(Y|X,A,Z,M)\big)]\nonumber \\
= & \mathbb{E}[\varphi_{2}(X,A,Z)S_{0}(X)]+\mathbb{E}[\varphi_{2}(X,A,Z)S_{0}(A|X)]+\mathbb{E}[\varphi_{2}(X,A,Z)S_{0}(Z|X,A)]\nonumber \\
 & +\mathbb{E}\big[\varphi_{2}(X,A,Z)\underbrace{\mathbb{E}[S_{0}(M|X,A,Z)|X,A,Z]}_{=0}\big]+\mathbb{E}\big[\varphi_{2}(X,A,Z)\underbrace{\mathbb{E}[S_{0}(Y|X,A,Z,M)|X,A,Z,M]}_{=0}\big]\nonumber \\
= & \mathbb{E}[\varphi_{2}(X,A,Z)S_{0}(X)]+\mathbb{E}[\varphi_{2}(X,A,Z)S_{0}(A|X)]+\mathbb{E}[\varphi_{2}(X,A,Z)S_{0}(Z|X,A)]\nonumber \\
= & \mathbb{E}\big[S_{0}(X)\underbrace{\mathbb{E}[\varphi_{2}(X,A,Z)|X,A]}_{=0}\big]+\mathbb{E}\big[S_{0}(A|X)\underbrace{\mathbb{E}[\varphi_{2}(X,A,Z)|X,A]}_{=0}\big]+\mathbb{E}[\varphi_{2}(X,A,Z)S_{0}(Z|X,A)]\nonumber \\
= & \mathbb{E}[\varphi_{2}(X,A,Z)S_{0}(Z|X,A)]\nonumber \\
= & \mathbb{E}\Big[\frac{\mathbb{I}(A=a)}{\Pr[A=a|X]}\big(\mathbb{E}[Y|X,A=a,Z,M=m]-\varphi_{1}(X)\big)S_{0}(Z|X,A)\Big]\nonumber \\
= & \mathbb{E}_{X}\Big\{\mathbb{E}\Big[\frac{\mathbb{I}(A=a)}{\Pr[A=a|X]}\big(\mathbb{E}[Y|X,A=a,Z,M=m]-\varphi_{1}(X)\big)S_{0}(Z|X,A)\Big\vert X,A=a\Big]\cdot\Pr[A=a|X]\Big\}\nonumber \\
= & \mathbb{E}_{X}\mathbb{E}\Big[\big(\mathbb{E}[Y|X,A=a,Z,M=m]-\varphi_{1}(X)\big)S_{0}(Z|X,A)\Big\vert X,A=a\Big]\nonumber \\
= & \mathbb{E}_{X}\mathbb{E}_{Z|X,A=a}\big[\mathbb{E}[Y|X,A=a,Z,M=m]S_{0}(Z|X,A)\big]\nonumber \\
= & \phi_{2}\label{eq:psi2}
\end{align}
where the fifth line follows from the fact that
\begin{align*}
\mathbb{E}[\varphi_{2}(X,A,Z)|X,A] & =\mathbb{E}[\frac{\mathbb{I}(A=a)}{\Pr[A=a|X]}\big(\mathbb{E}[Y|X,A=a,Z,M=m]-\varphi_{1}(X)\big)|X,A]\\
 & =\text{\ensuremath{\frac{\mathbb{I}(A=a)}{\Pr[A=a|X]}}}\big[\underbrace{\mathbb{E}_{Z|X,A}\mathbb{E}[Y|X,A=a,Z,M=m]}_{=\varphi_{1}(X)}-\varphi_{1}(X)\big]\\
 & =0.
\end{align*}
Third,
\begin{align}
 & \mathbb{E}[[\varphi_{3}(X,A,Z,M,Y)S_{0}(O)]\nonumber \\
= & \mathbb{E}[\varphi_{3}(X,A,Z,M,Y)\big(S_{0}(X)+S_{0}(A|X)+S_{0}(Z|X,A)+S_{0}(M|X,A,Z)+S_{0}(Y|X,A,Z,M)\big)]\nonumber \\
= & \mathbb{E}\big[\big(S_{0}(X)+S_{0}(A|X)+S_{0}(Z|X,A)+S_{0}(M|X,A,Z)\big)\underbrace{\mathbb{E}[\varphi_{3}(X,A,Z,M,Y)|X,A,Z,M]}_{=0}\big]\nonumber \\
 & +\mathbb{E}[\varphi_{3}(X,A,Z,M,Y)S_{0}(Y|X,A,Z,M)]\\
= & \mathbb{E}[\varphi_{3}(X,A,Z,M,Y)S_{0}(Y|X,A,Z,M)]\nonumber \\
= & \mathbb{E}\Big[\frac{\mathbb{I}(A=a)\mathbb{I}(M=m)YS_{0}(Y|X,A,Z,M)}{\Pr[A=a|X]\Pr[M=m|X,A=a,Z]}\Big]\nonumber \\
 & -\mathbb{E}\Big[\frac{\mathbb{I}(A=a)\mathbb{I}(M=m)\mathbb{E}[Y|X,A=a,Z,M=m]}{\Pr[A=a|X]\Pr[M=m|X,A=a,Z]}\underbrace{\mathbb{E}\big[S_{0}(Y|X,A,Z,M)|X,A,Z,M\big]}_{=0}\Big]\\
= & \mathbb{E}\Big\{\mathbb{E}\Big[\frac{\mathbb{I}(A=a)\mathbb{I}(M=m)YS_{0}(Y|X,A,Z,M)}{\Pr[A=a|X]\Pr[M=m|X,A=a,Z]}\Big\vert X,A,Z,M=m\Big]\Pr[M=m|X,A,Z]\Big\}\nonumber \\
= & \mathbb{E}_{X,A}\mathbb{E}\Big[\frac{\mathbb{I}(A=a)\Pr[M=m|X,A,Z]\mathbb{E}\big[YS_{0}(Y|X,A,Z,M)|X,A,Z,M=m\big]}{\Pr[A=a|X]\Pr[M=m|X,A=a,Z]}\Big\vert X,A\Big]\nonumber \\
= & \mathbb{E}_{X}\bigg\{\mathbb{E}\Big[\frac{\mathbb{I}(A=a)\Pr[M=m|X,A,Z]\mathbb{E}\big[YS_{0}(Y|X,A,Z,M)|X,A,Z,M=m\big]}{\Pr[A=a|X]\Pr[M=m|X,A=a,Z]}\Big\vert X,A=a\Big]\Pr[A=a|X]\bigg\}\nonumber \\
= & \mathbb{E}_{X}\bigg\{\mathbb{E}\Big[\mathbb{E}\Big[YS_{0}(Y|X,A,Z,M)\Big\vert X,A,Z,M=m\Big]\Big\vert X,A=a\Big]\bigg\}\nonumber \\
= & \phi_{3}\label{eq:psi3}
\end{align}
where the first equality follows from the fact that
\begin{align*}
 & \mathbb{E}[\varphi_{3}(X,A,Z,M,Y)|X,A,Z,M]\\
= & \mathbb{E}[\frac{\mathbb{I}(A=a)\mathbb{I}(M=m)\big(Y-\mathbb{E}[Y|X,A=a,Z,M=m]\big)}{\Pr[A=a|X]\Pr[M=m|X,A=a,Z]}|X,A,Z,M]\\
= & \frac{\mathbb{I}(A=a)\mathbb{I}(M=m)\big(\mathbb{E}[Y|X,A,Z,M]-\mathbb{E}[Y|X,A=a,Z,M=m]\big)}{\Pr[A=a|X]\Pr[M=m|X,A=a,Z]}\\
= & 0.
\end{align*}
Summing up equations (\ref{eq:psi1}-\ref{eq:psi3}), we have
\begin{align*}
 & \mathbb{E}[\varphi_{a,m}^{\textup{eff}}(O)S_{0}(O)]\\
= & \mathbb{E}[\big(\varphi_{1}(X)+\varphi_{2}(X,A,Z)+\varphi_{3}(X,A,Z,M,Y)-\psi_{am}\big)S_{0}(O)]\\
= & \mathbb{E}[\varphi_{1}(X)S_{0}(O)]+\mathbb{E}[\varphi_{2}(X,A,Z)S_{0}(O)]+\mathbb{E}[\varphi_{3}(X,A,Z,M,Y)S_{0}(O)]-\psi_{am}\cdot\mathbb{E}[S_{0}(O)]\\
= & \ensuremath{\phi_{1}}+\ensuremath{\phi_{2}+\phi_{3}-0}\\
= & \frac{\partial\psi_{am}(t)}{\partial t}\biggl\vert_{t=0}.
\end{align*}
One way to show the local efficiency of $\hat{\psi}_{am}^{\textup{tr}_{1}}$
is to verify that $\mathbb{E}[\varphi_{am}(O;\eta)]$ has a zero derivative
with respect to the nuisance functions $\eta=(\mu_{y}^{\textup{w}}(X,Z),\nu_{y}^{\textup{w}}(X),\pi_{a}^{\textup{w}}(X),\pi_{m}^{\textup{w}}(X,Z))$
at the truth. Suppose these nuisance functions are parameterized by
different components of a vector-valued parameter $\beta$, where
$\beta_{0}$ denotes the truth. We then have
\begin{align*}
\frac{\partial\varphi_{am}(O;\eta)}{\partial\beta}\Big\vert_{\beta=\beta_{0}}= & \frac{\mathbb{I}(A=a)}{\pi_{a}(X)}\big[1-\frac{\mathbb{I}(M=m)}{\pi_{m}(X,Z)}\big]\dot{\mu}_{y}^{w}(X,Z)\\
 & +\big[1-\frac{\mathbb{I}(A=a)}{\pi_{a}(X)}\big]\dot{\nu}_{y}^{w}(X)\\
 & -\frac{\mathbb{I}(A=a)}{\pi_{a}^{2}(X)}\big[\mu_{y}(X,Z)-\nu_{y}(X)+\frac{\mathbb{I}(M=m)}{\pi_{m}(X,Z)}\big(Y-\mu_{y}(X,Z)\big)\big]\dot{\pi}_{a}(X)\\
 & -\frac{\mathbb{I}(A=a)\mathbb{I}(M=m)}{\pi_{a}(X)\pi_{m}^{2}(X,Z)}\big[Y-\mu_{y}(X,Z)\big]\dot{\pi}_{m}(X,Z)
\end{align*}
where $\dot{\mu}_{y}^{w}(X,Z)$, $\dot{\nu}_{y}^{w}(X)$, $\dot{\pi}_{a}(X)$,
and $\dot{\pi}_{m}(X,Z)$ denote the derivatives of the corresponding
functions with respect to $\beta$ at $\beta_{0}$. It is easy to
verify that these components all have a zero mean. Thus $\mathbb{E}[\partial\varphi_{am}(O;\beta_{0})/\partial\beta]=0$,
implying that the influence function of $\hat{\psi}_{am}^{\textup{tr}_{1}}$
at $\mathcal{P}_{1}\cap\mathcal{P}_{3}$ is $\varphi_{a,m}^{\textup{eff}}(O)$.

Alternatively, we can also analyze the asymptotic expansion of $\hat{\psi}_{am}^{\textup{tr}_{1}}$
to establish weaker conditions for its semiparametric efficiency.
Denote $m(O;\eta)=\varphi_{am}(O;\eta)+\text{\ensuremath{\psi_{am}}}$,
we have 
\begin{align}
\hat{\psi}_{am}^{\textup{tr}_{1}} & =\mathbb{P}_{n}[m(O;\hat{\eta})]\nonumber \\
 & =\mathbb{P}_{n}[m(O;\eta)]+P[m(O;\hat{\eta})-m(O;\eta)]+(\mathbb{P}_{n}-P)[m(O;\hat{\eta})-m(O;\eta)],\label{eq:expansion}
\end{align}
where $Pg=\int gdP$ denotes the expectation of function $g(O)$ taken
at the truth. In equation \eqref{eq:expansion}, the first term can
be analyzed with the standard central limit theorem and has an asymptotic
variance of $\mathbb{E}[(\varphi_{am}(O;\eta))^{2}]$. The last term
is an empirical process term that will be $o_{p}(1/\sqrt{n})$ if
either the nuisance functions fall in a Donsker class or if cross-fitting
is used to induce independence between $\hat{\eta}$ and $O$ (\citealt{chernozhukov2018double}).
By rearranging terms, using the law of iterated expectations, and
applying the Cauchy-Schwartz inequality, we can rewrite the second
term as 
\begin{align*}
P[m(O;\hat{\eta})-m(O;\eta)] & =P\Big[\frac{(\hat{\pi}_{a}(X)-\pi_{a}(X))(\hat{\nu}_{y}(X)-\nu_{y}(X))}{\hat{\pi}_{a}(X)}\Big]\\
 & +P\Big[\frac{\mathbb{I}(A=a)(\hat{\pi}_{m}(X,Z)-\pi_{m}(X,Z))(\hat{\mu}_{y}(X,Z)-\mu_{y}(X,Z))}{\hat{\pi}_{a}(X)\hat{\pi}_{m}(X,Z)}\Big].\\
 & \leq C_{1}\Vert\hat{\pi}_{a}(X)-\pi_{a}(X)\Vert\cdot\Vert\hat{\nu}_{y}(X)-\nu_{y}(X)\Vert\\
 & +C_{2}\Vert\hat{\pi}_{m}(X,Z)-\pi_{m}(X,Z)\Vert\cdot\Vert\hat{\mu}_{y}(X,Z)-\mu_{y}(X,Z)\Vert,
\end{align*}
where $C_{1}$ and $C_{2}$ are on the order of $O_{p}(1)$, and $\Vert g\Vert=(\int g^{T}gdP)^{1/2}$.
The last line is due to the positivity assumption that $\pi_{a}(X)$
and $\pi_{m}(X,Z)$ are bounded away from zero. Thus the second term
in equation \eqref{eq:expansion} is asymptotically negligible if
$R_{n}(\hat{\pi}_{a})R_{n}(\hat{\nu}_{y})+R_{n}(\hat{\pi}_{m})R_{n}(\hat{\mu}_{y})=o(n^{-1/2})$,
where $R_{n}(\cdot)$ maps a nuisance function to its $L_{2}(P)$
convergence rate. This result implies that if all nuisance functions
are consistently estimated and converge at faster than $n^{1/4}$
rates, then $\hat{\psi}_{am}^{\textup{tr}_{1}}$ is semiparametric
efficient.

\section{More Details of the Simulation Study\label{sec:Simulation-Details}}

The variables $X,A,Z,M,Y$ in the simulation study are generated via
the following model:
\begin{align*}
(U_{XA},U_{XZ},U_{XM},U_{XY}) & \sim N(0,I_{4})\\
X & \sim N((U_{XA},U_{XZ},U_{XM},U_{XY})\beta_{X},1)\\
A & \sim\textup{Bernoulli}\big(\textup{logit}^{-1}[(1,U_{XA},|X|)\beta_{A}]\big)\\
Z & \sim N\big((1,U_{XZ},X,X^{2},A)\beta_{Z},1\big)\\
M & \sim\textup{Bernoulli}\big(\textup{logit}^{-1}[(1,U_{XM},X,X^{2},A,Z,XA,XZ)\beta_{M}]\big)\\
Y & \sim N\big((1,U_{XY},X,X^{2},A,Z,XZ,M,AM)\beta_{Y},1\big).
\end{align*}
The coefficients $\beta_{X},\beta_{A},\beta_{Z},\beta_{M},\beta_{Y}$
are generated using a set of uniform distributions with certain constraints
designed to create nontrivial degrees of model misspecification.

It can be shown that under the above model, the nuisance functions
$\mu_{y}(x,z)$, $\nu_{y}(x)$, $\pi_{a}(x)$, $\pi_{m}(x,z)$ can
be consistently estimated via the following GLMs:
\begin{align*}
\mathbb{E}[Y|X,A,Z,M] & =(1,X,X^{2},A,Z,XZ,M,AM)\theta;\quad\mu_{y}(X,Z)=\mathbb{E}[Y|X,A=a,Z,M=m]\\
\mathbb{E}[U|X,A] & =(1,X,X^{2},X^{3},A,XA)\eta;\quad\nu_{y}(X)=\mathbb{E}[U|X,A=a]\\
\pi_{a}(X) & =\textup{logit}^{-1}[(1,|X|)\alpha]\\
\pi_{m}(X,Z) & =\textup{logit}^{-1}[(1,X,X^{2},A,Z,XA,XZ)\gamma].
\end{align*}
Here $U$ is the outcome variable used to fit the model for $\nu_{y}(X)$,
as shown in equations (\ref{eq:nu1}-\ref{eq:nu3}). To demonstrate
the multiple robustness of the proposed estimators, we also fit a
misspecified model for each of the nuisance functions:
\begin{align*}
\mathbb{E}[Y|X,A,Z,M] & =(1,X,A,Z,M)\tilde{\theta};\quad\mu_{y}(X,Z)=\mathbb{E}[Y|X,A=a,Z,M=m]\\
\mathbb{E}[U|X,A] & =(1,X,A)\tilde{\eta};\quad\nu_{y}(X)=\mathbb{E}[U|X,A=a]\\
\pi_{a}(X) & =\textup{logit}^{-1}[(1,X)\tilde{\alpha}]\\
\pi_{m}(X,Z) & =\textup{logit}^{-1}[(1,X)\tilde{\gamma}].
\end{align*}
Each of the four cases described in Figure \ref{fig:simulation} reflects
a combination of estimated nuisance functions from these correctly
and incorrectly specified models. For example, for submodel $\mathcal{P}_{1}$,
we use correctly specified models for $\mu_{y}(x,z)$ and $\pi_{a}(x)$
and incorrectly specified models for $\nu_{y}(x)$ and $\pi_{m}(x,z)$
for all estimators.
\end{document}